\documentclass[preprint2]{aastex}

\usepackage{amsmath}
\usepackage{graphicx}
\usepackage{epsfig,psfrag,epic,eepic}
\usepackage{textcomp}
\usepackage{times}
\usepackage{longtable}

\slugcomment{For submission to the Astrophysical Journal Letters}

\shorttitle{Methane in the Atmosphere of GJ 504 b}
\shortauthors{Janson et al.}

\begin{document}

\title{Direct Imaging Detection of Methane in the Atmosphere of GJ 504 b}

\author{Markus Janson\altaffilmark{1,2}, 
Timothy~D. Brandt\altaffilmark{2}, 
Masayuki Kuzuhara\altaffilmark{3,4}, 
David~S. Spiegel\altaffilmark{5}, 
Christian Thalmann\altaffilmark{6}, 
Thayne Currie\altaffilmark{7}, 
Michael Bonnefoy\altaffilmark{8}, 
Neil Zimmerman\altaffilmark{8}, 
Satoko Sorahana\altaffilmark{9}, 
Takayuki Kotani\altaffilmark{4}, 
Joshua Schlieder\altaffilmark{8}, 
Jun Hashimoto\altaffilmark{4}, 
Tomoyuki Kudo\altaffilmark{10}, 
Nobuhiko Kusakabe\altaffilmark{4},
Lyu Abe\altaffilmark{11}, 
Wolfgang Brandner\altaffilmark{8}, 
Joseph~C. Carson\altaffilmark{12}, 
Sebastian Egner\altaffilmark{10}, 
Markus Feldt\altaffilmark{8}, 
Miwa Goto\altaffilmark{13}, 
Carol~A. Grady\altaffilmark{14,15}, 
Olivier Guyon\altaffilmark{10}, 
Yutaka Hayano\altaffilmark{10}, 
Masahiko Hayashi\altaffilmark{4}, 
Saeko Hayashi\altaffilmark{10}, 
Thomas Henning\altaffilmark{8}, 
Klaus~W. Hodapp\altaffilmark{16}, 
Miki Ishii\altaffilmark{10}, 
Masanori Iye\altaffilmark{4}, 
Ryo Kandori\altaffilmark{4}, 
Gillian~R. Knapp\altaffilmark{2}, 
Jungmi Kwon\altaffilmark{17}, 
Taro Matsuo\altaffilmark{18}, 
Michael~W. McElwain\altaffilmark{14}, 
Kyle Mede\altaffilmark{3}, 
Shoken Miyama\altaffilmark{19}, 
Jun-Ichi Morino\altaffilmark{4}, 
Amaya Moro-Mart\'in\altaffilmark{20}, 
Takao Nakagawa\altaffilmark{21}, 
Tetsuro Nishimura\altaffilmark{10}, 
Tae-Soo Pyo\altaffilmark{10}, 
Eugene Serabyn\altaffilmark{22}, 
Takuya Suenaga\altaffilmark{17}, 
Hiroshi Suto\altaffilmark{4}, 
Ryuji Suzuki\altaffilmark{23}, 
Yasuhiro Takahashi\altaffilmark{24}, 
Michihiro Takami\altaffilmark{25}, 
Naruhisa Takato\altaffilmark{10}, 
Hiroshi Terada\altaffilmark{10}, 
Daego Tomono\altaffilmark{10}, 
Edwin~L. Turner\altaffilmark{2,26}, 
Makoto Watanabe\altaffilmark{27}, 
John Wisniewski\altaffilmark{28}, 
Toru Yamada\altaffilmark{29}, 
Hideki Takami\altaffilmark{10}, 
Tomonori Usuda\altaffilmark{10}, 
Motohide Tamura\altaffilmark{4,17}
}

\altaffiltext{1}{Astrophysics Research Centre, Queen's University Belfast, University Road, BT7 1NN Belfast, Northern Ireland, UK; \texttt{m.janson@qub.ac.uk}}
\altaffiltext{2}{Department of Astrophysical Sciences, Princeton University, NJ 08544, USA}
\altaffiltext{3}{Department of Earth and Planetary Science, The University of Tokyo, 7-3-1 Hongo, Bunkyo-ku, Tokyo 113-0033, Japan}
\altaffiltext{4}{National Astronomical Observatory of Japan, 2-21-1 Osawa, Mitaka, Tokyo 181-8588, Japan}
\altaffiltext{5}{Astrophysics Department, Institute for Advanced Study, Princeton, NJ 08540, USA}
\altaffiltext{6}{Institute for Astronomy, ETH Zurich, Wolfgang-Pauli-Strasse 27, CH-8093 Zurich, Switzerland}
\altaffiltext{7}{Department of Astronomy and Astrophysics, University of Toronto, 50 St George Street, Toronto, ON M5S 3H8, Canada}
\altaffiltext{8}{Max Planck Institute for Astronomy, K\"onigstuhl 17, 69117 Heidelberg, Germany}
\altaffiltext{9}{Division of Particle and Astrophysical Science, Nagoya University, Furo-Cho, Chikusa, Nagoya 464-8602, Japan}
\altaffiltext{10}{Subaru Telescope, 650 North Aohoku Place, Hilo, HI 96720, USA}
\altaffiltext{11}{Laboratoire Lagrange, UMR7239, University of Nice-Sophia Antipolis, CNRS, Observatoire de la Cote d'Azur, 06300 Nice, France}
\altaffiltext{12}{Department of Physics and Astronomy, College of Charleston, 58 Coming Street, Charleston, SC 29424, USA}
\altaffiltext{13}{Universit\"ats-Sternwarte M\"unchen, Ludwig-Maximilians-Universit\"at, Scheinerstr. 1, 81679 Munich, Germany}
\altaffiltext{14}{Exoplanets and Stellar Astrophysics Laboratory, Code 667, Goddard Space Flight Center, Greenbelt, MD 2071, USA}
\altaffiltext{15}{Eureka Scientific, 2452 Delmer, Suite 100, Oakland CA 96002, USA}
\altaffiltext{16}{Institute for Astronomy, University of Hawai`i, 640 North A'ohoku Place, Hilo, HI 96720, USA}
\altaffiltext{17}{Department of Astronomical Science, Graduate University for Advanced Studies (Sokendai), Tokyo 181-8588, Japan}
\altaffiltext{18}{Department of Astronomy, Kyoto University, Kitsahirakawa-Oiwake-cho, Sakyo-ku, Kyoto, 606-8502, Japan}
\altaffiltext{19}{Office of the President, Hiroshima University, 1-3-2 Kagamiyama, Hagashi-Hiroshima, 739-8511, Japan}
\altaffiltext{20}{Department of Astrophysics, CAB (INTA-CSIC), Instituto Nacional de T\'ecnica Aerospacial, Torrej\'onde Ardoz, 28850, Madrid, Spain}
\altaffiltext{21}{Institute of Space and Astronautical Science, Japan Aerospace Exploration Agency, Sagamihara, 252-5210, Kanagawa, Japan}
\altaffiltext{22}{Jet Propulsion Laboratory, California Institute of Technology, Pasadena, CA 91109, USA}
\altaffiltext{23}{TMT Observatory Corporation, 1111 South Arroyo Parkway, Pasadena, CA 91105, USA}
\altaffiltext{24}{Department of Astronomy, The University of Tokyo, Hongo, Bunkyo-ku, Tokyo 113-0033, Japan}
\altaffiltext{25}{Institute of Astronomy and Astrophysics, Academia Sinica, P.O. Box 23-141, Taipei 106, Taiwan}
\altaffiltext{26}{Kavli Institute for the Physics and Mathematics of the Universe, The University of Tokyo, Kashiwa 277-8568, Japan}
\altaffiltext{27}{Department of Cosmosciences, Hokkaido University, Sapporo 060-0810, Japan}
\altaffiltext{28}{H.L. Dodge Department of Physics and Astronomy, University of Oklahoma, 440 W Brooks St Norman, OK 73019, USA}
\altaffiltext{29}{Astronomical Institute, Tohoku University, Aoba, Sendai 980-8578, Japan}

\begin{abstract}\noindent
Most exoplanets detected by direct imaging so far have been characterized by relatively hot ($\gtrsim$1000~K) and cloudy atmospheres. A surprising feature in some of their atmospheres has been a distinct lack of methane, possibly implying non-equilibrium chemistry. Recently, we reported the discovery of a planetary companion to the Sun-like star GJ 504 using Subaru/HiCIAO within the SEEDS survey. The planet is substantially colder ($<$600~K) than previously imaged planets, and has indications of fewer clouds, which implies that it represents a new class of planetary atmospheres with expected similarities to late T-type brown dwarfs in the same temperature range. If so, one might also expect the presence of significant methane absorption, which is characteristic of such objects. Here, we report the detection of deep methane absorption in the atmosphere of GJ 504 b, using the Spectral Differential Imaging mode of HiCIAO to distinguish the absorption feature around 1.6~$\mu$m. We also report updated $JHK$ photometry based on new $K_{\rm s}$-band data and a re-analysis of the existing data. The results support the notion that GJ 504 b has atmospheric properties distinct from other imaged exoplanets, and will become a useful reference object for future planets in the same temperature range.
\end{abstract}


\section{Introduction}
\label{s:introduction}

Several companions to stars in or near the planetary regime have been discovered through direct imaging in recent years \citep[e.g.][]{marois2010,lagrange2010,carson2013}, which has provided the opportunity for characterization of their atmospheres. The HR~8799 system in particular has been the subject of several studies using spectroscopy and multi-band photometry. These observations have provided indications of atmospheric properties such as non-equilibrium chemistry \citep[e.g.][]{janson2010,hinz2010,bowler2010} and thick clouds \citep[e.g.][]{barman2011,currie2011,skemer2012}. The latter implies that there may be a difference in cloud retention between $\sim$1000~K planets and more massive brown dwarfs in the same temperature range. Brown dwarfs undergo the so-called L/T transition \citep[e.g.][]{kirkpatrick2005} at these temperatures, which is thought to indicate a transition from cloudy (L-type) to clear (T-type) photopheres \citep[e.g.][]{leggett2000,chabrier2000,allard2001}. The apparent difference between planets and brown dwarfs in this regard can be theoretically justified through a dependence of the transitional temperature on surface gravity \citep{metchev2006,marley2012}.

Recently, we presented the discovery of GJ~504~b \citep{kuzuhara2013}, a planetary companion to a nearby G0-type star, made with direct imaging as part of the `Strategic Exploration of Exoplanets and Disks with Subaru' program \citep[SEEDS;][]{tamura2009}. With an age of $>$100~myr and a temperature of less than 600~K, GJ~504~b allows for study of a new class of planetary atmospheres, where the clouds have largely disappeared from the photosphere, more clearly revealing the molecular contents of the atmospheres. In this paper, we present new data in intermediate-band methane filters within $H$-band and re-analyze the broad-band photometry from \citet{kuzuhara2013} in order to acquire a uniformly reduced data set. In the following, we first describe the new observations that have been acquired, and the reduction procedure for the acquired data, as well as the re-reduction of the previous broad-band data. We then summarize the results, and compare them to a grid of models in order to constrain physical parameters, and finally we discuss the broader implications of our results. 

\section{Observations and Data Reduction}
\label{s:observations}

Here, we discuss the $J$ and $H$-band observations that have already been presented by \citet{kuzuhara2013}, new observations in $CH_4S$ and $CH_4L$, and a re-observation in $K_{\rm s}$-band. $CH_4S$ and $CH_4L$ are intermediate-band methane filters ranging from 1.486~$\mu$m to 1.628~$\mu$m for the former and from 1.643~$\mu$m to 1.788~$\mu$m for the latter. These filters are designed to capture both sides of the 1.6~$\mu$m methane bandhead expected in cool giant planet atmospheres under chemical equilibrium. If strong methane absorption is present, there should be significantly more flux in the $CH_4S$ band, which is outside of the methane absorption feature, than in the $CH_4L$ band, which is inside of it. The $CH_4S$ and $CH_4L$ images were acquired on 15 May 2012 using the HiCIAO camera \citep{suzuki2010} with the AO188 adaptive optics system \citep{hayano2008} in the 2-channel SDI (simultaneous spectral differential imaging) mode, in which the incoming beam is split into two beams using a Wollaston prism. One beam is passed through the $CH_4S$ filter and one through the $CH_4L$ filter, and the resulting two images are placed side by side on the detector. As with all the other observations of GJ~504, the Angular Differential Imaging \citep[ADI;][]{marois2006} mode was used, where the image rotator is used to keep the orientation of the telescope pupil fixed on the detector (pupil tracking). An exposure time of 20 seconds was used for each frame, without co-adding. During the continuous sequence of 106 minutes (including overheads), 219 frames were acquired with these settings, spanning a field rotation of 94.4$^{\rm o}$. In addition, non-saturated images were acquired interspersed with the ADI sequence (50 frames in total) for photometric referencing. These used 1.5~s exposures (without co-adding) and a neutral density (ND) filter. The ND filter has a transmission of 0.859\% across the broad $H$-band, but varies across wavelength such that the transmission is 0.714\% in $CH_4S$ and 0.929\% in $CH_4L$. The $K_{\rm s}$-band observation was acquired on 11 May 2012 in the same way as for the previous $K_{\rm s}$-band measurement of \citet{kuzuhara2013} from 28 Feb 2012, using HiCIAO in the ADI mode. Each co-added exposure consisted of 10 co-adds with individual integration times of 10.0~s. A total of 38 co-added exposures were acquired, spanning a field rotation of 68.1$^{\rm o}$. All observations used a circular occulting mask with a diameter of 0\farcs4.

Since the Wollaston prism splits up the beam based on polarization state, there is in principle a risk that either instrumental polarization or polarizated light from the planet itself could impact the apparent photometry in the SDI imaging. However, these effects are small (note that the planetary flux is emitted light rather than reflected light), particularly as the field rotation during the observation was more than 90$^{\rm o}$, which means that any polarization effect should be efficiently averaged out across the observation. Hence, we treat these effects as negligible in this analysis.

Data reduction was performed using the ACORNS-ADI pipeline in a uniform manner, described in \citet{brandt2013}. In addition to the $CH_4S$, $CH_4L$, and $K_{\rm s}$-band data, the $J$ and $H$-band data from \citet{kuzuhara2013} were re-reduced as well, since the discovery paper used a different pipeline and different reduction parameters and we wanted a uniform procedure. We also wished to examine whether uncertainties in the centroiding could have led to slightly higher photometric uncertainties than had previously been accounted for. Consequently, particular attention was committed to the registration procedure, which was applied after standard de-striping and flat fielding of the data. Errors in image centroiding and registration propagate into the derived companion photometry, and always result in a lower measured companion flux. An incorrect absolute centroid will dilute the companion flux in a de-rotated and collapsed image. The impact of this effect scales with the amount of field rotation during the observation. Errors in image registration reduce the measured companion photometry as shown in Figure 6 of \cite{brandt2013}; for example, root-mean-square errors of one HiCIAO pixel in both the horizontal and vertical directions reduce the measured near-IR photometry by $\sim$20\%.

The images of GJ 504 were nearly all taken with an opaque mask covering the central star, making registration and centroiding difficult: a positional shift by the star relative to the mask does not simply appear as a positional shift in the image on the detector. We therefore started out by not attempting to register the raw masked images, nor attempting to fit an absolute centroid directly. Instead, we initially assumed the frame-to-frame pointing stability to be perfect, estimated the centroid by eye, and then performed LOCI \citep[Locally Optimized Combination of Images;][]{lafreniere2007}. We then de-rotated the image sequence around many possible absolute centroids and co-added each resulting sequence. In this way, it becomes possible to find the centroid offset that maximizes GJ~504~b's signal-to-noise ratio in the final, reduced frame.  

Our method partially compensates for a real, stable drift of the stellar centroid, by also fitting for the drift speed and direction that maximizes the companion signal-to-noise ratio. Typical centroid drifts over a long observation are generally not more than 1--2 pixels \citep{brandt2013}, with a $\sim$2--3 pixel drift observed over a multi-hour observation at very high airmass \citep{thalmann2011}. We therefore expect the uncorrected bias from a stable drift to be much less than the 20\% that would have been introduced by positional fluctuations of 1 pixel, as mentioned above \citep{brandt2013}. Likewise, the actual frame-to-frame positional fluctuations in SEEDS images are small, $\lesssim$0.3 pixels \citep{brandt2013} and will have little effect on our photometry.

Major variations in atmospheric conditions and AO performance over the course of an observing run present another photometric difficulty. While it is possible to estimate the data quality from the diffraction pattern on masked frames, HiCIAO does not introduce an unsaturated central peak (as would a semitransparent mask) for a more precise calibration. Unsaturated photometric frames were sometimes taken only before and after an observing sequence and sometimes interspersed within a sequence. While most observations had stable conditions and consistent photometry across the sequences, the May 2011 $H$-band data presented a more difficult case. The unsaturated photometric frames taken before and after the sequence of masked science frames showed a remarkable uniformity in AO performance, and the science data themselves seemed to show no major AO failures or changes in observing conditions. However, using only the first half of the sequence yielded a planet flux $\sim$50\% higher than that derived from the latter half of the sequence. This discrepancy corresponds to a 2$\sigma$ significance level. None of the other data sets showed any differences of this magnitude for companion photometry in subgroups of science frames.

All frames were subjected to a LOCI-based PSF (Point Spread Function) modeling and subtraction with standard ACORNS-ADI parameters ($N_{\rm FWHM} = 0.6$, $N_{\rm Area} = 200$). We also tried varying the parameters, but since the final image is not strongly dominated by PSF residuals at the location of the companion, this had a negligible effect. All the PSF-subtracted data were de-rotated to a common field orientation and co-added to form the final image of each filter. The primary flux (rescaled from the neutral density filter images in the case of $CH_4S$ and $CH_4L$) as well as the secondary flux were calculated in circular apertures with a diameter equal to the 75\% of the FWHM. Partial subtraction as a function of angular separation imposed by LOCI was accounted for when determining the contrasts.

\section{Results and Discussion}
\label{s:results}

\begin{figure*}[p]
\centering
\includegraphics[width=16cm]{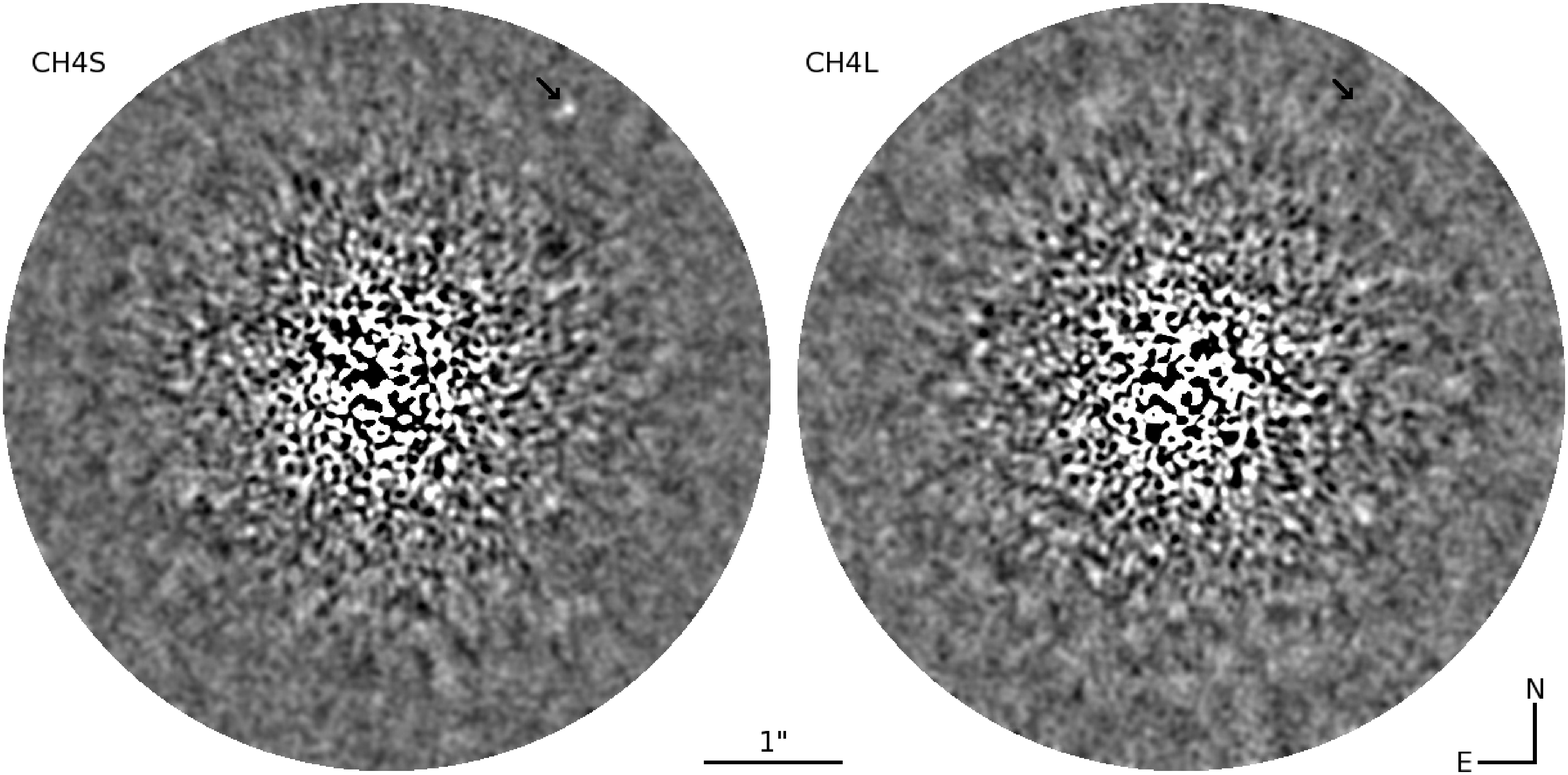}
\caption{Image showing the methane absorption in GJ~504~b. Left panel: The $CH_4S$ filter, outside of the methane absorption band. Right panel: The $CH_4L$ filter, inside of the methane absorption band. The planet is bright in $CH_4S$ but completely disappears in $CH_4L$, implying strong methane absorption.}
\label{f:ch4im}
\end{figure*}

The final $CH_4S$ and $CH_4L$ images resulting from the reduction are shown in Fig. \ref{f:ch4im}, and the photometric values are shown in Table \ref{t:photometry}. It can be immediately seen from the images that the planet is bright in $CH_4S$ but disappears in $CH_4L$. This corresponds exactly to the behavior expected from an atmosphere with methane absorption, given that $CH_4L$ encompasses the methane aborption feature, whereas $CH_4S$ represents the continuum flux. Apparent magnitudes for the companion were calculated as the measured contrast to the primary plus the magnitude of the primary in the given filter. Since there are no literature values for the primary's $CH_4S$ and $CH_4L$ magnitudes, we translated the $H$-band magnitude into these bands by relating the stellar flux to the Vega flux in the relevant passbands, assuming blackbody radiation with a $T_{\rm eff}$ of 6234~K for GJ~504 (G0V) and 9602~K for Vega (A0V). As expected, the corrections were very small, with $CH_4S-H = 0.02$~mag and $CH_4L-H = -0.01$~mag. Apparent magnitudes were translated into absolute magnitudes using the parallax distance of 17.56$\pm$0.08~pc from Hipparcos \citep{vanleeuwen2007}.

\begin{table}[p]
\caption{Photometry of GJ 504 b.}
\label{t:photometry}
\centering
\begin{tabular}{lcc}
\hline
\hline
Band & GJ 504 A & GJ 504 b \\
\hline
$J$	&	4.11$\pm$0.01~mag	&	19.76$\pm$0.10~mag	\\
$H$	&	3.86$\pm$0.01~mag	&	19.99$\pm$0.10~mag\tablenotemark{a}	\\
$K_{\rm s}$	&	3.81$\pm$0.01~mag	&	19.38$\pm$0.11~mag	\\
$CH_4S$	&	3.88$\pm$0.01~mag	&	19.59$\pm$0.12~mag	\\
$CH_4L$	&	3.85$\pm$0.01~mag	&	$>$20.62~mag (3$\sigma$)	\\
\hline
\end{tabular}
\tablenotetext{a}{There is an apparent systematic difference between the first and second \\
halves of the data set, see text for discussion.}
\end{table}

\begin{figure}[p]
\centering
\includegraphics[width=8cm]{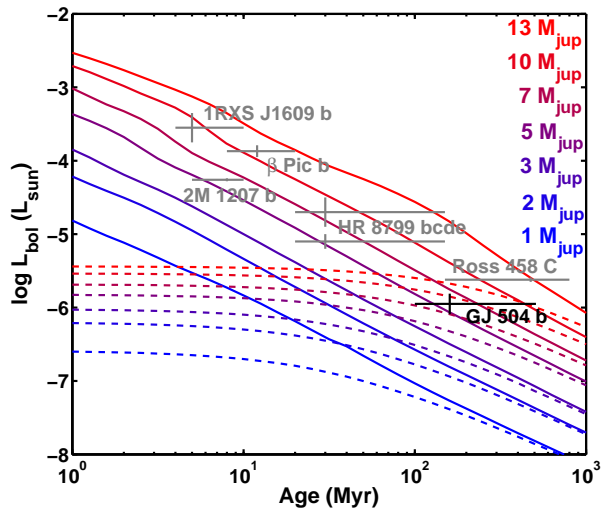}
\caption{Evolution of luminosity with age as a function of mass for the coldest (dashed lines) and hottest (solid lines) initial conditions of typical warm-start models \citep{spiegel2012}. The relatively old age of GJ~504~b means that it is only weakly sensitive to the initial conditions. Other objects included in the figure are $\beta$~Pic~b \citep{lagrange2010}, 1RXS~J1609~b \citep{lafreniere2010}, 2M1207~b \citep{chauvin2005} and the HR~8799 planets \citep{marois2008,marois2010}.} 
\label{f:cooling}
\end{figure}

\begin{figure}[p]
\centering
\includegraphics[width=8cm]{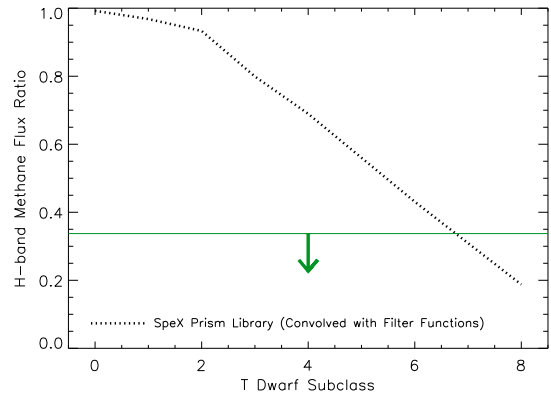}
\caption{A comparison of the GJ~504~b methane flux ratio limit to the spectral type sequence of field brown dwarfs. The green horizontal line marks the measured limit of GJ~504~b. The dotted line represents our calculated indices from the SpeX prism library.}
\label{f:ch4color}
\end{figure}

The updated $J$ and $H$ photometry is $\sim$0.1~mag brighter than the previous photometry. This is consistent within the error bars, but we adopt the new photometry as it more accurately represents the full set of flux losses possible in the reduction procedure. Our $K_{\rm s}$-band photometry based on the newer (May 2012) data point alone is $\sim$0.3~mag fainter than previously, which is again consistent given the larger scatter in the Feb 2012 data. The $K_{\rm s}$-band photometry presented here is based on a weighted mean of the two epochs\footnote{We note here that the newer (May 2013) $K_{\rm s}$-band image has in fact already been plotted in \citet{kuzuhara2013}.}. We note that the changes in $JHK$ values of this order correspond to changes of a small fraction of 1~$M_{\rm Jup}$ in inferred mass ($\sim$0.1--0.2~$M_{\rm Jup}$) regardless of the choice of mass-luminosity model, hence there is no significant change to the mass estimation presented in \citet{kuzuhara2013}. On this topic, it is also worth noting that the uncertainties in physical quantities from these observational uncertainties are strongly dominated by model uncertainties, as could also be seen in \citet{kuzuhara2013}. For instance, comparing the old and the new measurements within the context of the \citet{spiegel2012} models, the differences in inferred temperature and luminosity are 1\% and 6\%, respectively. By contrast, the difference for equal measurements between the \citet{baraffe2003} models and the \citet{spiegel2012} models are 14\% in temperature and 51\% in luminosity. This is simply because the \citet{spiegel2012} models generally predict fainter magnitudes for a given effective temperature and luminosity. We use the dominating span between the models as the uncertainties when we plot the luminosity and temperature in Figs. \ref{f:cooling} and \ref{f:ch4color}. 

Using the full range in luminosity, we show in Fig. \ref{f:cooling} that the mass estimate remains well constrained to the planetary mass range, irrespective of initial conditions. The agreement between the photometry of our new $K_{\rm s}$-band observation with the previous one also demonstrates the general robustness of the photometry. However, one issue that should be kept in mind is that the $H$-band observation displays a puzzling behavior, where the companion appears to have a 2$\sigma$ different brightness in the first and second halves of the data set, apparently irrespective of calibration issues. It may be that something (e.g., temporary cloud coverage) occurred during some part of the observation such that it should not be trusted. However, there are no visual or quantitative clues in the data to support such an assessment. In general, it should be recalled that the difference of the two halves of the data set could simply be a statistical fluctuation, as 2$\sigma$ is not a statistically unreasonable deviation. We note that, as was mentioned by \citep{kuzuhara2013}, the average $H$-band photometry of the May 2011 epoch is consistent with the photometry from the March 2011 discovery epoch, which may support the statistical fluctuation interpretation. Future observations will be necessary to conclusively settle this issue.

\begin{figure}[p]
\centering
\includegraphics[width=8cm]{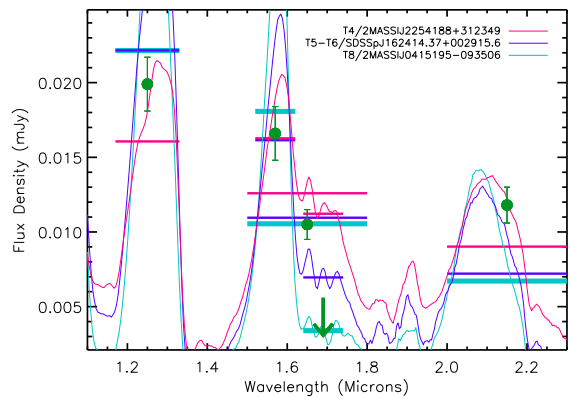}
\caption{Comparison of GJ~504~b to the calculated magnitudes of field brown dwarfs from mesured spectra \citep{mclean2003} in the Spex prism library. Only late T-type objects can reproduce the deep methane absorption seen in GJ~504~b.}
\label{f:empirical}
\end{figure}

With the $CH_4S$ and $CH_4L$ observations taken simultaneously, the methane detection is robust to any photometric calibration uncertainties. The absorption is deep and appears consistent with late T-dwarfs in the field. In order to quantify this, we calculated spectral indices from brown dwarfs in the SpeX prism library\footnote{Maintained by Adam Burgasser at http://pono.ucsd.edu/
$\sim$adam/browndwarfs/spexprism} using $CH_4S$ and $CH_4L$ filter functions. A comparison to GJ~504 ~b is shown in Fig. \ref{f:ch4color}. It can be seen that the methane absorption of the planet is consistent with spectral types later than T6. To further check how GJ~504~b compares to empirical spectra, we also calculated $JHK$ photometry from the SpeX library in addition to $CH_4S$ and $CH_4L$, and derived $\chi ^2$ after applying a best-fit scaling factor and penalizing spectra predicting a $CH_4L$ detection \citep{currie2011}. Examples from the comparison are shown in Fig. \ref{f:empirical}. This analysis again suggests a late spectral type, with T8 providing a better fit than earlier spectral types, although the best fit still is not impressive, due to the excess brightness at $K$-band of GJ~504~b relative to field dwarfs already noted in \citet{kuzuhara2013}.

Methane has otherwise remained elusive in exoplanetary atmospheres thus far. While indications of methane have been suggested for the transiting planets HD~189733~b and WASP-12~b, these interpretations are ambiguous \citep[e.g.][]{swain2008,gibson2011,mandell2011,stevenson2013}. Also for the case of directly imaged planets, there has been a trend of CO/CO$_2$  being the dominant carbon-bearing species, rather than CH$_4$ as is expected under chemical equilibrium \citep[e.g.][and references above]{lodders2002,konopacky2013,lee2013}. Methane absorption has been identified in the atmosphere of HR~8799~b \citep{barman2011}, though at modest strength. A stronger case of methane absorption has been reported in Ross~458~C \citep{burgasser2010}, which is a very wide ($\sim$1000~AU) companion to an M-dwarf binary with an inferred mass near the deuterium-burning limit. In \citet{kuzuhara2013}, we noted that the colors of GJ~504~b are similar to those of GJ~758~B \citep{thalmann2009}, which is a similarly cold brown dwarf companion to a Sun-like star. Our current detection of methane further underlines this similarity, as GJ~758~B also features strong methane absorption \citep{janson2011}. This strengthens the indications that GJ~504~b belongs to a different class of atmospheres than most of the previously imaged planets. The latter share similarities with L-type or L/T transition brown dwarfs, while GJ~504~b appears to have more in common with the T spectral type range. However, like GJ~758~B, it is anomalously bright in the $K$-band range, which could be a signature of high metallicity or possibly low surface gravity. Indeed, this feature of high $K$-band brightness relative to field brown dwarfs of similar temperature is a feature that is also present in other objects with indications of very low gravity, such as CFBDSIR~J214947.2-040308.9 \citep{delorme2012} and Ross~458~C \citep[e.g.]{goldman2010,burningham2011}.

The similarity of GJ~504~b to late T-type objects, implying relatively little cloud coverage \citep[although remnant clouds cannot be excluded, see e.g.][]{burgasser2010,morley2012}, is particularly interesting in the context that thick clouds have been inferred even for HR~8799~b,  whose temperature may be as low as $\sim$800~K \citep[e.g.][]{barman2011}, significantly colder than the temperature of $\sim$1400~K at which brown dwarfs lose their thick clouds \citep[e.g.][]{allard2001}. This implies that the transition from a more to a less cloudy regime may occur somewhere in the temperature range of 600--800~K for objects in the planetary mass range, and is consistent with the predictions of \citet{marley2012}. Future detections of planets in this intermediate range would be critical to test this hypothesis. The necessary quantity of such objects may be delivered by near-future facilities such as GPI \citep{macintosh2008}, SPHERE \citep{beuzit2008}, or CHARIS \citep{peters2012}. Similarly, it will be useful to study planetary mass companions to very low-mass stars and brown dwarfs \citep[e.g.][]{chauvin2005}, or even free-floating objects in this mass range \citep[e.g.][]{delorme2012} in order to test for possible similarities or distinctions between these different populations, and thus to clarify mutual trends with surface gravity, or divergent trends from formation scenarios. 

\acknowledgements
Part of this work was supported by NASA through Hubble Fellowship grant HF-51290.01 awarded by the Space Telescope Science Institute, which is operated by the Association of Universities for Research in Astronomy, Inc., for NASA, under contract NAS 5-26555, and through NSF award 1009203. We thank the referee, Adam Burgasser, for his very useful comments. This study made use of the CDS services SIMBAD and VizieR, as well as the SAO/NASA ADS service.  The study has benefitted from the SpeX Prism Spectral Libraries at \url{http://www.browndwarfs.org/spexprism}.

\clearpage

\end{document}